\documentclass[aps,superscriptaddress,showpacs,nofootinbib,twocolumn]{revtex4}
\usepackage{bm}

\newcommand{\be}{\begin{equation}}
\newcommand{\ee}{\end{equation}}
\newcommand{\bea}{\begin{eqnarray}}
\newcommand{\eea}{\end{eqnarray}}
\begin{document}

\title{Gauge symmetry and Slavnov-Taylor identities for randomly stirred fluids}

\author{Arjun Berera}
\email{ab@ph.ed.ac.uk}
\affiliation{
School of Physics,
University of Edinburgh, Edinburgh EH9 3JZ, Great Britain}
\author{David Hochberg}
\email{hochbergd@inta.es}
\affiliation{Centro de
Astrobiolog\'{\i}a (CSIC-INTA), Ctra. Ajalvir Km. 4, 28850
Torrej\'{o}n de Ardoz, Madrid, Spain}

\begin{abstract}
The path integral for randomly forced incompressible fluids is shown to
have an underlying Becchi-Rouet-Stora (BRS) symmetry as a consequence of
Galilean invariance. This symmetry must be respected
to have a consistent generating functional, free from both an overall infinite
factor and spurious relations amongst correlation functions.
We present a procedure for respecting this BRS symmetry, akin to gauge
fixing in quantum field theory.
Relations are derived between correlation functions of this
gauge fixed, BRS symmetric theory, analogous
to the Slavnov-Taylor identities of quantum field theory.

\bigskip

In Press Physical Review Letters, 2007

\end{abstract}

\date{\today}

\maketitle

To be consistent with Newtonian physics, the description of fluid dynamics and
turbulence must be the same in all inertial reference frames.
For this reason, the role that Galilean invariance plays in the dynamics of a fluid
system governed by the Navier-Stokes equation has been
a major focus of study for decades \cite{FNS,DeDomMar}.
The Galilean invariance of the Navier-Stokes equation
plays an important part in the physical aspects of turbulence \cite{McComb}, in
practical aspects of turbulence modelling \cite{Pope} and even in fluid animation
and computer graphics applications \cite{Shah}.
Among the symmetries of the Navier-Stokes equation, 
Galilean invariance is by far the most
studied in theoretical approaches to turbulence \cite{FNS,DeDomMar,Spez,Teo,Poly,Mou}.
The physical constraint of Galilean invariance must be taken into account in the
modelling of subgrid-scale stresses in large eddy simulations \cite{Spez}, in models of
multi-component turbulence \cite{Mou}, in nonperturbative renormalization group analyses
of fully developed turbulence \cite{Toma}, and in probability density
transport equations \cite{Tong}. Much of the interest has dealt with the infinite number of
exact relations between different correlation functions implied
by Galilean invariance, which are akin to the Ward-Takahashi identities of quantum field
theory \cite{Ramond}.  The most well known of these exact identities relates the vertex and response
functions \cite{DeDomMar}.  Based on this relation, various inferences have been
made about the nonrenormalization of the advective or inertial term
in the Navier-Stokes equation \cite{FNS,Teo,Toma,Mou,DeDomMar,Mc,BH}.

In this Letter we observe that the Galilean invariant generating functional
for stirred fluids \cite{DeDom,DeDomMar,Eyink}, from which all the Ward identities
are derived, possesses an overall infinite factor,
which must first be extracted before this functional
is well-defined.  We draw an analogy with abelian gauge
theory in quantum field theory, whereby the Galilean invariance
is associated with gauge invariance, and so the infinity
arises from integrating the fluid velocity over gauge equivalent copies (i.e., over
an infinite number of independent inertial reference
frames) of the same theory.  To remove the infinite factor in
the Navier-Stokes generating functional, the Fadeev-Popov procedure
of quantum field theory \cite{fp} is used to ``gauge fix'' the theory,
that is, to fix the theory to a single inertial reference
frame. The gauge fixing results in the theory no longer appearing
Galilean invariant. However, we demonstrate that this gauge-fixed
theory possesses an underlying Becchi-Rouet-Stora
(BRS) symmetry \cite{BRS,Ramond,ZJ}, which when invoked,
restores the Galilean invariance.

This work provides a new insight into the structure
of the Navier-Stokes equation.  However, it is not simply a new
way to express the same theory.  We show that the
infinite factor associated with the standard
dynamic functional has associated with it, much more dire
consequences, in that there exist spurious relations amongst
correlation functions.  Explicit examples of such spurious relations
are given.  We then demonstrate that the gauge-fixing
BRS procedure implemented here not only removes the
infinite factor, but also eliminates these spurious relations,
thereby rendering a well defined generating functional for
this theory.  The BRS invariant functional
derived in this Letter is the consistent and correct
functional for the Navier-Stokes theory, that must be used to obtain
reliable results.

The standard dynamic generating functional for a stirred incompressible
fluid has been extensively studied for many years \cite{DeDom,Mou,Teo,Toma,DeDomMar,Eyink,BH}.
It is based on the path integral approach \cite{DeDom,Jan,Phythian,Jensen} to
classical statistical dynamics \cite{MSR} and is given by
\begin{equation}\label{Z}
Z = \int [D\mathbf{V}][D\bm{\sigma}] \exp \{
-S[\mathbf{V}, \bm{\sigma}] + \int d\mathbf{k}d\omega (\mathbf{J} \cdot \mathbf{V}
+ \bm{\Sigma} \cdot \bm{\sigma}) \},
\end{equation}
where the action
\begin{eqnarray}\label{classk}
& & S[\mathbf{V}, \bm{\sigma}] = \frac{1}{2}\int d\mathbf{k} d\omega \,\,
\mathbf{\sigma}_i(\mathbf{-k},- \omega)D_{ij}(-\mathbf{k}) \mathbf{
\sigma}_j(\mathbf{k},\omega) -
\nonumber  \\
&  & i\int d\mathbf{k} d\omega\,\, \sigma_{\alpha}(\mathbf{-k},-\omega)
\left[ \big(-i \omega + \nu k^2\big) V_{\alpha}(\mathbf{k},\omega) -
\right. \nonumber \\
& & \left. M_{\alpha \beta \gamma}(\mathbf{k})\int d\mathbf{j} d\omega_1
V_{\beta}(\mathbf{k - j},\omega - \omega_1) V_{\gamma}(\mathbf{
j}\,,\omega_1) \right].
\end{eqnarray}
The fluid velocity is $\mathbf{V}$, the conjugate field is $\bm{\sigma}$; $\mathbf{J}$ and
$\bm{\Sigma}$ are $\mathbf{k}$ and $\omega$ dependent sources.
$D_{ij}(-{\bf k})$ is the only non-vanishing cumulant of the random force,
$M_{\alpha \beta \gamma} (\mathbf{k}) = (2i)^{-1}\{ k_{\beta}
P_{\alpha \gamma}(\mathbf{k}) + k_{\gamma} P_{\alpha \beta}(\mathbf{k})\}$ and
$P_{\alpha \beta}({\bf k}) = \delta_{\alpha \beta} -
\frac{k_{\alpha} k_{\beta}}{|\mathbf{k}|^2}$,
and $\nu$ is the bare fluid viscosity.
Throughout this Letter we will work in wavevector-frequency space.

Consider a second primed reference frame moving with respect to the
unprimed frame at constant velocity
$\mathbf{c} = c {\hat {\bf n}}$ where ${\hat {\bf n}}$ is
a unit vector in the direction of motion.
The relations between wavevector, and the frequency of an event
in the two frames
and the transformations
of the velocity and conjugate fields are given by:
\begin{eqnarray}\label{kgal1}\mathbf{k}' = \mathbf{k} , & & \
\omega' = \omega - \mathbf{c \cdot k},\\
\label{kgal3} \mathbf{V}(\mathbf{k},\omega) &=&
\mathbf{V}'(\mathbf{k}, \omega - \mathbf{c \cdot k}) +
\mathbf{c}\,\delta^3(\mathbf{k}) \delta({\omega}),\\
\label{kgal4} \bm{\sigma}(\mathbf{k},\omega) &=& \bm{\sigma^\prime}
(\mathbf{k},\omega - \mathbf{c \cdot k}).
\end{eqnarray}
The action $S$ and functional measure are invariant under this Galilean transformation.
The Fadeev-Popov procedure of quantum field theory \cite{Ramond,fp}
is now applied to show that for ${\bf J} = \bm{\Sigma} = {\bf 0}$,
the functional in
Eq. (\ref{Z}) has an overall infinite factor.
For this, expressing the velocity zero mode as
$\mathbf{V}(\mathbf{0},0) = {\bf V}_0 \delta^3({\mathbf{0}})\delta(0)$ \cite{footnote1},
first insert the factor of
unity $1 = \int d{\bf b}\, \delta^3( {\bf V}_0 - {\bf b} )$, into Eq. (\ref{Z}),
to give
\begin{equation}\label{Zkomega}
Z = \int [D{\bf V}][D \bm{\sigma}]
\int d{\bf b}\, \delta^3( {\bf V}_0 - {\bf b} )
\exp \{ -S[{\bf V}, \bm{\sigma}]\}.
\end{equation}
The observation is now made that
the integrand in Eq. (\ref{Zkomega})
is independent of ${\bf b}$.
To see this, consider changing from ${\bf b} \rightarrow {\bf b}+ {\bf a}$.
Then perform a Galilean transformation with boost velocity
${\bf c} = {\bf a}$ in Eqs. (\ref{kgal1}) - (\ref{kgal4}).
The action and measure are invariant to this transformation and
$\delta^3( {\bf V_0} - ({\bf b} + {\bf a}) ) \rightarrow \delta^3({\bf V}_0 - {\bf b} )$,
thus restoring the integrand to its original form.
Due to the independence of the integrand on the value of
$\bf b$, the $d {\bf b}$ integral can completely factor out as an overall
infinity as
\begin{equation}\label{Zkomegafac}
Z = \left(\int d{\bf b} \right) \int [D{\bf V}][D \bm{\sigma}]
\delta^3({\bf V}_0 - {\bf b} )
\exp \{ -S[{\bf V}, \bm{\sigma}] \}.
\end{equation}
The first term in brackets on the right-hand side
is the infinite factor we originally sought to isolate. At this stage one could simply
eliminate this integral since it is just an overall,
albeit infinite, normalization factor.
In this case different choices of ${\bf b}$, correspond
to expressing the theory in different inertial reference frames.
On the other hand, since the integrand in
Eq. (\ref{Zkomega}) is independent of ${\bf b}$, we can also
insert any function $G({\bf b})$ into the integrand, for example
to render the ${\bf b}$ integration finite.
One convenient choice is,
$G({\bf b}) = \exp(-\frac{1}{2 \xi}{\bf b} \cdot {\bf b})$, where
$\xi > 0$.
Inserting this into Eq. (\ref{Zkomegafac}) and now performing
the $d {\bf b}$ integration, leaves the gauge-fixed (GF) functional
\begin{equation}\label{Zkomegagf}
Z^{\prime}_{GF} = \int [D{\bf V}][D \bm{\sigma}]
\exp \{ -S[{\bf V}, \bm{\sigma}] -
\frac{1}{2 \xi} {\bf V}_0 \cdot {\bf V}_0 \}.
\end{equation}
In our analogy to gauge theories, all these different
possibilities for $G({\bf b})$ above would be called gauge choices.

Problems also arise if one attempts to compute any
correlation function using Eq. (\ref{Z}). For example, consider
$\langle V_{\alpha}({\bf k},\omega) V_{\beta}({\bf k}',\omega')\rangle$.
Using the Galilean invariance of the action,
this can be written as
\begin{eqnarray}\label{Zdecomp}
&\langle& V_{\alpha}(\mathbf{k},\omega) V_{\beta}(\mathbf{k}',\omega')\rangle =
\int [D{\bf {\tilde V}}][D \bm{\sigma}] \exp \{ -S[{\bf {\tilde V}}, \bm{\sigma}]\}
\nonumber \\
&\times& \int [D{\bf {\tilde V}_c}]
V_{\alpha}({\bf k},\omega) V_{\beta}({\bf k}',\omega')/Z[{\bf 0}, {\bf 0}] ,
\end{eqnarray}
where the $[D{\bf {\tilde V}}]$ integration
is over all velocity field configurations that are not related by Galilean
invariance, and for each such field configuration the
$[D{\bf {\tilde V}_c}]$ integration is over all
field configurations related to the particular field by a Galilean
transformation. The expression Eq. (\ref{Zdecomp}) shows that using the
ungauge-fixed functional $Z$, Eq. (\ref{Z}), the
calculation of a correlation function sums over configurations from all
Galilean-related reference frames, which in general
will lead to an infinite, ill-defined quantity.
This same situation would hold for a correlation
function comprised of any combination
of fields. Moreover, since these infinities are not just simple terms
that factor out, it means they can not be removed from
the generating functional by simply dividing as
$Z[{\bf J}, \bm{\Sigma}]/Z[{\bf 0}, {\bf 0}]$.

The gauge-fixing term in Eq.(\ref{Zkomegagf}) manifestly breaks the Galilean
invariance of the action. We next demonstrate that this gauge fixed
theory possesses a symmetry
akin to the BRS symmetry of quantum field theory \cite{BRS,Ramond,ZJ},
which when recognized, \textit{restores}
the Galilean  invariance of the gauge fixed theory.
We can multiply the generating functional
$Z^{\prime}_{GF}$ in Eq.({\ref{Zkomegagf}) by the constant
$\int d \bm{\eta} d \bm{\eta}^* \exp\{i \bm{\eta}^* \cdot \bm{\eta}\}$
where $\bm{\eta}$ and $\bm{\eta}^*$ are constant complex conjugate
Grassmann vectors $\{\eta_i, \eta^*_j\}=0$, $\bm{\eta}^2 = \bm{\eta}^{*2} = 0$,
and are independent of {\bf k} and $\omega$.
For the ``$V_0^2$-gauge" of
Eq. (\ref{Zkomegagf})
consider the following (infinitesimal) BRS transformation
\begin{eqnarray}\label{brs1}
\delta_{\rm BRS} {\bf V}_0 & = & -c {\zeta} (\bm{\eta}^* + \bm{\eta}), \\
\label{brs2}
\delta_{\rm BRS} \bm{\eta}  =  -\frac{ i}{\xi} {\bf V}_0 c {\zeta}, & &
\delta_{\rm BRS} \bm{\eta}^*  =  +\frac{ i}{\xi} {\bf V}_0 c {\zeta},
\end{eqnarray}
where $c$ is a constant with dimensions of velocity,
$\zeta$ is a real Grassmann parameter (independent of \textbf{k} and $\omega$)
while all other modes of the velocity ${\bf V}({\bf k}, \omega)$
and $\bm{\sigma}({\bf k}, \omega)$ transform under $\delta_{\rm BRS}$ as implied by
Eqs. (\ref{kgal1}) - (\ref{kgal4}),
except we replace the velocity boost ${\bf c}$ by $\mathbf{c}\rightarrow
c {\zeta} (\bm{\eta}^* + \bm{\eta})$.
The gauge-fixed action
\begin{equation}\label{GFaction}
S_{GF}[\mathbf{V},\bm{\sigma},\bm{\eta},\bm{\eta}^*] = S[\mathbf{V},\bm{\sigma}]
+ \frac{1}{2 \xi} {\bf V}_0 \cdot {\bf V}_0 - i \bm{\eta}^* \cdot \bm{\eta}\ ,
\end{equation}
is BRS invariant: $\delta_{\rm BRS}S_{GF}[\mathbf{V},\bm{\sigma},\bm{\eta},\bm{\eta}^*] = 0$.
In the gauge fixed functional
\begin{eqnarray}
Z_{GF} & = &
\int [D{\bf V}][D \bm{\sigma}]d\bm{\eta} d\bm{\eta}^*
\exp \{ -S_{GF}[{\bf V}, \bm{\sigma},\bm{\eta},\bm{\eta}^*]
\nonumber \\
& + & \int d\mathbf{k}d\omega ({\bf J} \cdot {\bf V}
+ \bm{\Sigma} \cdot \bm{\sigma}) + \bm{\theta^*} \cdot \bm{\eta}
+ \bm{\theta} \cdot \bm{\eta^*} \},
\label{Zgf}
\end{eqnarray}
where $\bm{\theta}$ and $\bm{\theta^*}$ are constant
Grassman vector source terms,
it is this BRS invariance that replaces the Galilean transformation
and leads to relations amongst correlation functions,
which in analogy with quantum field theory, will be called
the Slavnov-Taylor identities \cite{Ramond,ZJ}.

To derive these, start with the above gauge-fixed functional
$Z_{GF}[\mathbf{J},\bm{\Sigma},\bm{\theta},\bm{\theta}^*]$
and displace all fields by the infinitesimal BRS transformation
Eqs.(\ref{kgal1}-\ref{kgal4}), with $\mathbf{c}\rightarrow
c{\zeta} (\bm{\eta}^* + \bm{\eta})$
and Eqs.(\ref{brs1},\ref{brs2}). Since the measure and the
gauge-fixed action $S_{GF}$ are BRS-invariant,
only the source terms will be displaced.
Moreover, as $\zeta^2 = 0$, we can easily
expand the exponential. The gauge-fixed functional transforms as
$Z_{GF} \rightarrow Z_{GF} + \delta_{BRS}Z_{GF}$,
where  $\delta_{BRS}Z_{GF} = 0$
can be written as follows \cite{footnote1}:
\begin{eqnarray}\label{deltaBRS}
[\frac{i}{\xi}(\bm{\theta}^* -
\bm{\theta})\cdot\frac{\delta}{\delta {\mathbf{J_0}}}+
(\frac{\partial}{\partial \theta_j}
+\frac{\partial}{\partial \theta_j^*}){\hat O}_{j}]
Z_{GF}=0,
\label{WTIgf}
\end{eqnarray}
where
\begin{eqnarray}
{\hat O}_{j}&=&
\int d\mathbf{k} d\omega \, \Big(
J_m(-)k_j \frac{\partial}{\partial \omega}\frac{\delta}{\delta J_m(-)}
\nonumber\\
&+&\Sigma_m(-)k_j\frac{\partial}{\partial \omega}\frac{\delta}{\delta \Sigma_m(-)}
-J_j(-)\delta(\mathbf{k})\delta(\omega) \Big).
\label{OWT}
\end{eqnarray}
The shorthand $(-)$ stands for $(-\mathbf{k},-\omega)$.

We now show that the gauge fixed BRS symmetric
theory Eq. (\ref{GFaction}) and the resulting 
Slavnov-Taylor identities in Eq. (\ref{WTIgf})
are not simply a different way
to write the known Ward identities \cite{DeDomMar} of this theory.
First we demonstrate that the standard generating functional
for the Navier Stokes equation Eqs. (\ref{Z}) and (\ref{classk}),
in the absence of gauge fixing,
contains spurious relations amongst
correlation functions.  To see this, return
to Eqs. (\ref{Z},\ref{classk})
and perform the Galilean transformations
Eqs. (\ref{kgal1}-\ref{kgal4}).
Since the measure and action $S$ are invariant under this
particular change of variables, it leads to the relation
\begin{eqnarray}
\label{spurious}
&& \langle \exp \int d\mathbf{k} d\omega \,
\{\mathbf{J}(-)\cdot\mathbf{V}(\mathbf{k},\omega)
+
\bm{\Sigma}(-)\cdot\bm{\sigma}(\mathbf{k},\omega)
\}\rangle \nonumber \\
&=& \langle \exp \int d\mathbf{k} d\omega \,
\{\mathbf{J}(-)\cdot\mathbf{V}(\mathbf{k},\omega +
\mathbf{c}
\cdot\mathbf{k})
\nonumber\\
&-&\mathbf{c}\cdot\mathbf{J}(-)\delta^3(\mathbf{k})\delta(\omega)
+\bm{\Sigma}(-)\cdot\bm{\sigma}(\mathbf{k},\omega
+ \mathbf{c} \cdot\mathbf{k})\}\rangle ,
\end{eqnarray}
where the averaging on both sides is with respect to
the same measure and action $S$ in Eq. (\ref{classk}). 
The above relation implies an infinite number
of spurious relations.  For example, acting with
one derivative $\delta / \delta J_i(-{\bf k}, \omega)$ on both
sides and setting ${\bf J}=\bm{\Sigma}= {\bf 0}$, it gives
$\langle {\bf V}_i({\bf k},\omega) \rangle
= \langle {\bf V}_i({\bf k},\omega + {\bf c}
\cdot {\bf k}) - {\bf c}_i\delta^3({\bf k})\delta(\omega)\rangle$.
Integrating over an infinitesimal neighborhood
near ${\bf k} = {\bf 0}$, $\omega=0$, it leads to the relation
${\bf c}_i = 0$,
which is meaningless, since ${\bf c}$ is an arbitrary real vector
which we are free to choose.
Likewise, taking two derivative
$\delta / \delta J_i(-{\bf k}_1, \omega_1) \delta / \delta J_j(-{\bf k}_2, \omega_2)$ of
Eq. (\ref{spurious}) when integrated over an infinitesimal
neighborhood near ${\bf k}_{1,2}={\bf 0}$,
${\omega}_{1,2}=0$ leads to the relation
$-c_j\Delta\mathbf{k}_1\Delta\omega_1\langle
\mathbf{V}_i(\mathbf{0},0)\rangle  -
c_i\Delta\mathbf{k}_2\Delta\omega_2\langle
\mathbf{V}_j(\mathbf{0},0)\rangle + c_ic_j = 0$,
which again requires ${\bf c} = {\bf 0}$, which is senseless.
In a similar fashion, any number of further derivatives
of ${\bf J}$ and/or $\bm{\Sigma}$ fields will lead
to an infinite number of spurious relations.
These examples show that at order ${\bf c}$ and all higher orders,
the generating functional Eq. (\ref{Z}) contains an infinite number
of spurious relations.  
Although many papers \cite{Teo,Mou,Toma,BH,Eyink}
have studied the generating functional Eq. (\ref{Z}),
to our knowledge this
Letter is the first to reveal this inherent inconsistency.

At $O(c)$, an alternative way to reveal spurious relations is
to work directly with the Ward identity expression.
For the standard generating functional,
the standard Ward identity relation found in the references \cite{Teo,Mou, Toma,BH} is
${\hat O}_j Z = 0$, where ${\hat O}_j$
is the operator in Eq. (\ref{OWT})
and $Z$ is the generating functional Eq. ({\ref{Z}).
Consider
$\delta/\delta {\bf J}_i(-{\bf k},-\omega) {\hat O}_j|_{{\bf J} = \bm{\Sigma} = {\bf 0}} Z  = 0$,
this leads to the relation
$\langle k_{j} \frac{\partial}{\partial \omega} {\bf V}_i({\bf k},\omega)
- \delta_{ij}\delta^3({\bf k})\delta(\omega) \rangle = 0$.
Contract this with $\delta_{ij}$, then fluid incompressibility
implies that $\delta^3({\bf k})\delta(\omega)=0$,
which is certainly false for $\mathbf{k} = \mathbf{0}$ and $\omega = 0$.
However, now we can exhibit an infinite number of such
spurious relations all at $O(c)$.  For example
if two derivatives are taken
$\delta/\delta {\bf J}_i(-{\bf k}_1,-\omega_1) \delta/
\delta {\bf J}_m(-{\bf k},-\omega){\hat O}_j |_{{\bf J} = \bm{\Sigma} = {\bf 0}}Z = 0$,
then after contraction, fluid incompressibility implies the relation
$\delta^3({\bf k}_1) \delta(\omega_1) \langle{\bf V}_m(\mathbf{k},\omega)\rangle =0$,
but the mean fluid velocity $\langle {\bf V}({\bf k}, \omega) \rangle$
can be generally nonzero for an appropriate noise force.
So, this relation is also spurious.
Moreover if one more derivative is taken with with
respect to ${\bf J}$ or $\bm{\Sigma}$, irrespective of the
properties of the noise force, the resulting relation would
be spurious.  And similarly, any additional number of such derivatives
will lead to an infinite number of spurious relations 
in which they will wrongly imply higher order
velocity correlation functions are vanishing.

In contrast, for the case of our gauge-fixed generating functional
Eq. (\ref{Zgf}), all spurious relations have been eliminated.
At all orders higher than O($c$), this follows elegantly from
the Grassmann property that $\zeta^2 = 0$.  And at O($c$)
the first nontrivial relation is obtained by differentiating,
Eq. (\ref{deltaBRS}) by
$\delta/\delta {\bf J_0}_l \delta/\delta \theta_j$ to
give
\begin{equation}
\frac{i}{\xi} \langle V_{0l} V_{0j} \rangle
+ \langle {{\eta}_j}^* {\eta}_l \rangle =0 ,
\label{spurfixed}
\end{equation}
which can be checked is correct with a similar relation
found by replacing the differentiation by $\theta_j$
with $\theta_j^*$.  
Further $O(c)$ relations can be
obtained by taking more ${\bf J}$ and/or more
${\bf \theta}$ or ${\bf \theta^*}$ derivatives, but due to
factorization, these
are all proportional to the fundamental 
identity Eq. (\ref{spurfixed}), and so the GF-BRS functional
Eq. (\ref{Zgf}) is completely free of spurious relations.

These considerations support our statement that the generating
functional Eq. (\ref{Zgf})
is the correct and consistent one for
this theory, and the correct expression of the Galilean invariance
is the Slavnov-Taylor identities Eq. (\ref{WTIgf}), found in this
Letter.

Up to now, we have been working with the generating functional $Z_{GF}$
Eq. (\ref{Zgf}),
but all our considerations can be applied to
the generator of connected diagrams $W = \ln Z_{GF}$ and
to the one-particle irreducible graphs $\Gamma$.
Introducing the generating functional of one particle
irreducible functions (i.e., the effective action)
\begin{eqnarray}
\Gamma[\mathbf{V}_{cl},\bm{\sigma}_{cl},\bm{\eta}_{cl},
\bm{\eta}_{cl}^*] &=& -W[\mathbf{J},
\bm{\Sigma},\bm{\theta},\bm{\theta}^*] + \bm{\theta}^*\cdot \bm{\eta}_{cl}
+ \bm{\theta}\cdot\bm{\eta}^*_{cl}\nonumber \\
&+&
\int d\mathbf{k}d\omega \,
(\mathbf{J}\cdot\mathbf{V}_{cl} + \bm{\Sigma}\cdot\bm{\sigma}_{cl}),
\end{eqnarray}
we see that $J_k = \frac{\delta \Gamma}{\delta V_k^{cl}},\,\,
\Sigma_k = \frac{\delta \Gamma}{\delta \sigma_k^{cl}},\,\,
\theta_j = \frac{\partial \Gamma}{\partial {\eta^*}^{cl}_j},\,\,
\theta^*_j = \frac{\partial \Gamma}{\partial {\eta}^{cl}_j}.$

The noise-averaged fields in the presence of the sources are denoted $\mathbf{V}_{cl} =
\langle \mathbf{V} \rangle_{sources}$, etc.
In terms of $\Gamma$ we can immediately write down the identity Eq.(\ref{WTIgf}) in the form
\begin{eqnarray}\label{WTI2}
&&{}\,\, 0 = \frac{i}{\xi}\mathbf{V}^{cl}_0\cdot\frac{\delta \Gamma}{\partial \bm{\eta}_{cl}} -
\frac{i}{\xi}\mathbf{V}^{cl}_0\cdot\frac{\delta \Gamma}{\partial \bm{\eta}^*_{cl}}\\
&+&
(\eta_{cl} + \eta^*_{cl})_j\int d\mathbf{k} d\omega \,
\Big(k_j \frac{\partial V^{cl}_m(\mathbf{k},\omega)}{\partial \omega}
\frac{\delta \Gamma}
{\delta V^{cl}_m(\mathbf{k},\omega)}\nonumber \\
&+& k_j \frac{\partial \sigma^{cl}_m(\mathbf{k},\omega)}{\partial \omega}
\frac{\delta \Gamma}
{\delta \sigma^{cl}_m(\mathbf{k},\omega)} - \delta(\mathbf{k})\delta(\omega)\frac{\delta \Gamma}
{\delta V^{cl}_j(\mathbf{k},\omega)}\Big).\nonumber
\end{eqnarray}
We now apply this formula to the problem at hand. The dependence of $\Gamma$ on
$\bm{\eta}$ and $\bm{\eta}^*$ is simple since these constant Grassmann vector
fields do not interact nor do they couple to the velocity or conjugate fields.
Thus, we can immediately write
$\Gamma[\mathbf{V}_{cl},\bm{\sigma}_{cl},\bm{\eta}_{cl},\bm{\eta}_{cl}^*] =
-i \bm{\eta}_{cl}^* \cdot \bm{\eta}_{cl} + \Gamma[\mathbf{V}_{cl},\bm{\sigma}_{cl}]$,
where $\Gamma[\mathbf{V}_{cl},\bm{\sigma}_{cl}]$ does not depend on either
$\bm{\eta}_{cl}$ or $\bm{\eta}_{cl}^*$, with the first few terms
written in \cite{Taylor}.

The pertinent identity we seek is then obtained by inserting $\Gamma$
into Eq.(\ref{WTI2}), and then differentiating this
with respect to
$\partial/{\partial \eta^{cl}_j}\, \delta/\delta V^{cl}_l(\mathbf{k},\omega)
\delta/\delta \sigma^{cl}_n(-\mathbf{k},-\omega)$ followed by setting
$\bm{\eta}^{cl} = \bm{\eta}^*_{cl} = \mathbf{V}^{cl}$ = $\bm{\sigma}^{cl}= 0$.
Note that the terms in Eq.(\ref{WTI2}) depending on the
gauge parameter $\xi$ do not contribute, and that we end up with
identity \cite{Teo,Mou,Toma,BH}
\begin{equation}\label{WTIk2}
-k_m\frac{\partial }{\partial \omega}
\Gamma_{ln}^{(1,1)}(-\mathbf{k},-\omega;\mathbf{k},\omega) = \Gamma_{mln}^{(2,1)}
(\mathbf{0},0; -\mathbf{k},-\Omega;\mathbf{k},\omega) .
\end{equation}
This well known identity, has been explicitly verified
up to one-loop order \cite{FNS,Teo}.  The GF-BRS theory preserves this
identity as it should, but this theory is moreover free of
all spurious relations.  In addition, the origin of this
identity is seen through the GF-BRS approach
as arising in the limit of vanishing mean translational velocity,
but completely independent of the properties of
the fluctuating velocity.
This conclusion, consistent with the assertions in \cite{Mc,BH},
is seen here in a systematic, mathematically consistent, derivation.

The implications and consequences of Galilean invariance for Navier-Stokes
fluid dynamics have been outstanding questions for some time now
\cite{FNS,DeDomMar,Spez,
Teo,Mou,Poly,Toma,Tong,Mc,BH}. The formulation of the
Navier Stokes equation subject to random forcing  as a classical stochastic
field theory
\cite{DeDom,Mou,Teo,Toma,DeDomMar,Eyink,BH} opens up the way to apply the
systematic
techniques originally developed for quantum fields \cite{ZJ}.
This Letter has shown that a field theoretic approach provides clear
understanding of
the Galilean symmetry through its close relation to gauge fixing and BRS
symmetry;
the constraint singling out a unique reference frame 
leads to a global symmetry
with anticommuting parameters first used in quantum gauge field theory
\cite{BRS}.

In summary, we have discovered a new symmetry
in Galilean invariant classical stochastic field theories,
which provides a powerful new way to
think about
the analysis of these systems.
In this Letter we focused on the Navier-Stokes equation but
in broader terms our analysis applies also to other Galilean
invariant theories such as
the  KPZ equation \cite{KPZ}, magnetohydrodynamics
and the Burgers equation.
An immediate and important consequence of our result
is that, within the presently understood generating functionals for
these theories, there are spurious relations
which our gauge fixing procedure corrects.  
Although the treatment in this Letter throughout assumed a
stochastic noise force, formally our treatment is also applicable
when the noise force is absent.
Our work, apart
from now providing the mathematically consistent expression
for functionals of Galilean invariant stochastic
field theories, would also be of importance in
any numerical/simulation calculations of these theories.

We thank W. D. McComb
for useful discussions.
Support was provided to A.B. by UK PPARC and
D.H. by CSIC-INTA (Spain).


\begin{thebibliography}{99}
\bibitem{footnote1} $\mathbf{V}(\mathbf{0},0)$ and $\mathbf{V_0}$ are
related by a space-time volume factor. In a bounded box 
of spatial side $L$ and temporal side $T$,
this factor goes as $L^3T$. For an unbounded box, this factor becomes $
\int d^3x\,\int dt= \delta^3(\mathbf{0})\delta(0)$. 
The source $\mathbf{J_0} = \delta^3(\mathbf{0})
\delta(0)\mathbf{J}(\mathbf{0},0)$.
\end{thebibliography}
\end{document}